\documentclass{he_symp}
\usepackage{psfig,graphicx,epsfig}
\usepackage{color}
\usepackage{amsmath,amssymb,epic,eepic,array}
\begin{document}
\renewcommand{\FirstPageOfPaper }{ 162}\renewcommand{\LastPageOfPaper }{ 166}

\title{Emission Mechanisms in High Energy Pulsars: From Gamma Rays to Infrared}
\author{Andr\'e R. Crusius--W\"atzel \and Harald Lesch}
\institute{Institut f\"ur Astronomie und Astrophysik der Universit\"at
M\"unchen, Scheinerstr. 1, 81679 M\"unchen, Germany}
\authorrunning{Crusius-W\"atzel and Lesch}
\titlerunning{Emission Mechanisms in High Energy Pulsars}
\maketitle

\begin{abstract}
We present a model for the gamma-ray emission of pulsars in terms
of curvature radiation by highly relativistic particles. It is
shown that the injection of power-law primary particles from an outer gap
and subsequent cooling by curvature radiation losses reproduces the
high energy spectrum and the luminosity of pulsars. As a
result a spectral break is to be expected at a photon energy of
$\sim 1\thinspace{\rm GeV}$. This value does not depend on the
surface magnetic field strength or on the period of the neutron
star, but only involves a geometric factor. The predicted
change of the spectral index by 1 explains in particular the
spectral shape of PSR B1706--44.
We find that according to this model the
luminosity of pulsars in the high energy band varies according to
$L_{\gamma}\propto B_0/P^{5/3}$, which is in good agreement with the
observations.

A model for the infrared, optical and soft X-ray emission
of pulsar is also presented. It is based on anisotropic synchrotron
emission by relativistic particles in an
outer gap scenario with a single energy distribution
$N(\gamma)\propto \gamma^{-2}$.
It is shown that this synchrotron model is able to reproduce
the spectral shape between the infrared and soft X-rays and also
the corresponding luminosities for the Crab pulsar.
In particular, the long standing problem of the rapid spectral decline towards
infrared frequencies is understandable as emission at very small
pitch angles from low energy particles with $\gamma\ga 10$.
It is also shown that the scaling of the synchrotron model explains
the observed correlation between the X-ray luminosity and the spin-down
luminosity of the neutron star $L_{\rm x} \sim 10^{-3} {\dot E}$.
\end{abstract}

\section{Gamma-Ray Emission of Pulsars}

There are now at least seven spin-powered pulsars known to be
emitting pulsed gamma radiation. Detailed spectral information has been
obtained by observations with the {\sl Compton Gamma Ray Observatory}
in the energy range of up to 10 GeV.
Several of these pulsars have a break in the spectrum
at about $1-2\thinspace{\rm GeV}$. This is also the
energy where the gamma-ray pulsars emit most of the spectral
power. The gamma ray luminosity is correlated to the square root
of the spin-down power of these pulsars (with some scatter).
For specific informations about the properties of individual
gamma ray pulsars we refer to Thompson et al. (1996), Fierro et al. (1998),
Thompson et al. (1999), and Kanbach (2001).

The models proposed for high-energy pulsars can be devided into
polar cap models (e.g., Daugherty and Harding 1996; Sturner
and Dermer 1994, Usov and Melrose 1996) and outer gap models
(e.g., Romani and Yadigaroglu 1995; Romani 1996). In the first group
the acceleration of particles and the gamma ray production
is done in the open field line region above the magnetic pole
of the neutron star. In the
second group this takes place in the vacuum gaps between
the neutral line and the last open field line in the
outer magnetosphere, near the light cylinder. Depending on the
distribution of the electric field parallel to the magnetic field
along and across the gap, a power-law energy distribution of
primary particles can be expected that emit gamma radiation.
Here we discuss the effects
of subsequent cooling by curvature radiation.

\subsection{Primary Particles, and Particle Energy Spectrum}

Particles entering the starvation gap will be accelerated
by the strong electric field along the magnetic field.
The accelerating electric field given in Cheng, Ho,
and Ruderman (1986) can be rewritten as
\begin{equation}
E_0=\delta^2B\ ,
\end{equation}
where $\delta\sim 0.3$ is the ratio of the gap thickness to the
light cylinder radius.
The particles gain energy at a rate $P_{\rm gain}=ceE_0$.
The energy of these primary particles is limited by curvature radiation losses, that depend on the bending of the field lines.
The equilibrium between gains and losses determines the maximum energy.
The curvature radius $R_{\rm c}$ of the last open
field line is of the order of the
light cylinder radius $R_{\rm lc}=c/\Omega$, with $\Omega=2\pi/P$,
and increases towards the neutral line. This becomes even more important for
a (nearly) perpendicular rotator. By setting $R_{\rm c}=kR_{\rm lc}$,
with $k\sim 1$,
the power emitted by a single particle is then
\begin{equation} \label{curve-power}
P_{\rm cr}=\frac{2e^2c}{3R_{\rm c}^2}\gamma^4
=\frac{2e^2}{3c}k^{-2}\Omega^2\gamma^4\ .
\end{equation}
The typical frequency emitted is
\begin{equation}
\nu_{\rm c}=\frac{3c}{4\pi R_{\rm c}}\gamma^3
=\frac{3}{4\pi}k^{-1}\Omega\gamma^3\ .
\end{equation}
Equating $P_{\rm gain}$ with $P_{\rm loss}=P_{\rm cr}$,
this results in an equilibrium Lorentz factor for the primaries of
\begin{equation} \label{gamma-max}
\gamma_{\rm max}=4.6\times 10^7\ (k\delta)^{1/2}\kappa^{-3/4}
\left(\frac{B_{0,12}}{P_{-1}}\right)^{1/4}\ ,
\end{equation}
where $\kappa\sim 0.5$ is the ratio of the inner starting radius of
the gap to the light cylinder radius.

The emission of gamma ray photons by the primaries leads to a pair
creation cascade. These pairs, if produced inside the gap,
will also be accelerated, start to shield the electric potential
in the gap. As their density increases and becomes of the
order of the Goldreich-Julian density $n_{\rm GJ}=\Omega B/2\pi ec$
this finally closes the gap. The outward moving particles are no
longer accelerated and only loose energy via curvature radiation.

The pairs generated outside the gap with a finite pitch angle
due to aberration effects will emit synchrotron radiation
that can be seen in the MeV, X-ray and optical bands (Romani, 1996;
Crusius-W\"atzel, Kunzl, and Lesch, 2001).
The electric field available for the acceleration of particles
depends on the details of the cascade process, which leads to a quasi
stationary state where the electric field is a function of the location
along and across the gap. Although attempts have been made to understand
the cascade and outer gap closure, no solid model exists. Therefore
we just assume a power-law energy distribution of the accelerated
particles,
\begin{equation} \label{injection}
Q(\gamma)\propto\gamma^{-s}\ ,\ \gamma_{\rm min}<\gamma<\gamma_{\rm max}\ ,
\end{equation}
with an index $s$, that must be determined from the observations,
rather than from theory in the light of the considerations given above.
Since $\nu\propto\gamma^3$ the ratio
$\gamma_{\rm max}/\gamma_{\rm min}$ needs to be only about 5
to cover a frequency range of more that two decades.

\subsection{Curvature Radiation Cooling}

The timescale for curvature cooling of relativistic electrons or positrons
near the light cylinder of the pulsar magnetosphere is given by
\begin{equation}
\tau=\frac{\gamma}{\dot\gamma}=4.1\times
10^{-2}k^2P_{-1}^2\gamma_7^{-3}\ {\rm s}\ ,
\end{equation}
where $\dot\gamma=P_{\rm cr}/mc^2$ and equation
(\ref{curve-power}) has been used. We estimate the Lorentz factor
$\gamma_{\rm b}$ above which we expect a spectral break by setting
the cooling timescale equal to the time needed for the particle to
move out of the magnetosphere into the wind region, $t_{\rm esc}=fR_{\rm
lc}/c=f/\Omega$, where $f\sim 0.1$ is the fraction of the light
cylinder radius, corresponding to the outer gap region in which
the particles are not accelerated anymore but only radiate.
From this consideration we find that
cooling will be important for particles with $\gamma>\gamma_{\rm
b}$, where
\begin{equation} \label{break}
\gamma_{\rm b}=\left(\frac{3mc^3k^2}{2e^2f\Omega}\right)^{1/3}
=2.9\times 10^7k^{2/3}f_{-1}^{-1/3}P_{-1}^{1/3}\ .
\end{equation}
The frequency at which the spectral break occurs is then
\begin{equation} \label{frequency}
\nu_{\rm b}=\frac{3\Omega}{4\pi k}\gamma_{\rm b}^3
=\frac{9mc^3}{8\pi e^2}\cdot\frac{k}{f}\ ,
\end{equation}
which corresponds to a frequency and photon energy of
$\nu_{\rm b}=3.8\times 10^{23}(k/f_{-1}){\rm Hz}$ and
\begin{equation}
E_{\rm b}=h\nu_{\rm b}=1.6\thinspace{\rm GeV}\frac{k}{f_{-1}}\ ,
\end{equation}
respectively,
depending only on the geometrical factors $k$ and
$f$. In particular, {\bf the break energy is independent of
the magnetic field strength and period}.

\subsection{Gamma-Ray Spectrum}

For frequencies below the break $\nu_{\rm b}$ the spectrum can be
calculated from the particle energy distribution according to
equation (\ref{injection}). By using
the monochromatic approximation for the spectral power of a single
lepton, $P_{\nu}(\gamma)=P_{\rm cr}(\gamma)\delta[\nu-\nu_{\rm c}(\gamma)]$,
it follows that
\begin{eqnarray}
I_{\nu}&\propto&\int d\gamma\thinspace Q(\gamma)P_{\nu}(\gamma) \\
&\propto&\int d\gamma\thinspace
\gamma^{-s}\gamma^4\left|\frac{d(\nu-\nu_{\rm c})}{d\gamma}\right|^{-1}
\delta(\gamma-\gamma_0)\ , \nonumber
\end{eqnarray}
where $\gamma_0\propto \nu^{1/3}$ is the Lorentz factor at which the
argument of the $\delta$-function becomes zero. We then find that
\begin{equation}
I_{\nu}\propto\nu^{-(s-2)/3}\ \ (\nu<\nu_{\rm b})\ .
\end{equation}
This corresponds to a photon energy index of
$\phi(\nu<\nu_{\rm b})=-(s+1)/3$, where
$N_{\rm E}\propto E^{\phi}$.
The Crab, Geminga, and Vela pulsar have photon indices
-2, -1.4, and -1.6, respectively, below the spectral break
(Fierro, et al., 1998). This then gives the particle energy indices
$s$ of the primaries below the break energy as 5, 3.2, and 3.8,
respectively.

For frequencies above the break we first have to find the distribution
function modified by radiation losses. Assuming that a spectrum $Q(\gamma)$
is injected at a rate ${\dot Q}(\gamma)=Q(\gamma)/t_{\rm esc}$, the cooled
spectrum is given by the following expression
\begin{equation}
N(\gamma)=\frac{1}{|\dot\gamma|}\int_\gamma^{\gamma_{\rm max}}
d\gamma'\thinspace {\dot Q}(\gamma')\ ,
\end{equation}
from which we find $N(\gamma)\propto \gamma^{-(s+3)}$. By using
the monochromatic approximation again, we finally get
\begin{equation}
I_{\nu}\propto\nu^{-(s+1)/3}\ \ (\nu>\nu_{\rm b})\ ,
\end{equation}
corresponding to a photon index of $\phi(\nu>\nu_{\rm b})=-(s+4)/3$.
Therefore the spectral index of the gamma radiation changes by
\begin{equation}
\Delta\phi=-1
\end{equation}
at the break frequency (eqn. \ref{frequency}) in the direction
of increasing frequencies.

The pulsar PSR B1706--44 shows exactly this behavior
(Thompson et al., 1996). The gamma ray photon spectrum changes by
$\Delta\phi=-1$, from $\phi=-1.25$ below 1~GeV to $\phi=-2.25$
above 1~GeV, with the spectrum extending for more than a decade in
energy, both, below and above the break. From this we find
$s=2.75$ for this pulsar.

The emitted power per logarithmic frequency band scales as
$\nu I_{\nu}\propto\nu^{(5-s)/3}$ below the break and
$\nu I_{\nu}\propto\nu^{(2-s)/3}$ above the break.
For $2<s<5$ the gamma ray energy distribution thus is rising below and
is falling above the break energy. The maximum power is then emitted
at the break energy.

The highest frequencies are produced by the primary particles
that feel the vacuum potential. The spectrum of the radiation
emitted by them cuts off exponentially above
\begin{eqnarray} \label{numax}
\nu_{\rm max}&=&\frac{3\Omega\gamma_{\rm max}^3}{4\pi k} \\
&=&1.5\times 10^{24}\delta^{3/2}k^{1/2}\kappa^{-9/4}B_{0,12}^{3/4}P_{-1}^{-7/4}
\thinspace {\rm Hz}\ .\nonumber
\end{eqnarray}
The change in the spectrum is dominated by cooling only if the exponential
cutoff is at higher energies than the break, $\nu_{\rm max}>\nu_{\rm b}$.
The cooled spectrum is nicely seen in PSR B1706--44, where the cutoff must be
above 10~GeV. The situation is less clear for the other gamma ray pulsars,
where the energy range between the break and the cutoff is either smaller
or not existent. It can be expected that with the GLAST mission more
pulsars with a clear break will be found and also that a smaller range of the
cooled spectrum can be detected.

\subsection{Luminosity Estimates for the Gamma-Ray Emission}

We now estimate the power $L_{\gamma}$ emitted by pulsars
in the gamma-ray region at the break frequency very roughly
as the product of the volume,
the Goldreich-Julian number density near the light cylinder
$n_{\rm GJ}\propto B_0/P^4$,
and the radiation power of a single particle
$P_{\rm cr}\propto\gamma^4/P^2$,
\begin{equation}
L_{\gamma}=Vn_{\rm GJ}P_{\rm cr}\ ,
\end{equation}
where the volume is assumed to be a shell near the light cylinder
with $V=4\pi f R_{\rm lc}^3\propto P^3$. From this we find
\begin{equation} \label{luminosity}
L_{\gamma}=4.2\times 10^{34}\thinspace\frac{k^{2/3}}{\kappa^{3}f_{-1}^{1/3}}
\frac{B_{0,12}}{P_{-1}^{5/3}}\ {\rm erg\thinspace s^{-1}}\ ,
\end{equation}
where the break Lorentz factor from equation (\ref{break}) has been used
in the curvature radiation power (\ref{curve-power}). Since the spin-down
luminosity $\dot E$ of a pulsar is proportional to $B_0^2/P^4$,
the gamma luminosity thus scales according to
$L_{\gamma}\propto {\dot E}^{1/2}P^{1/3}$. The observed high-energy
luminosities of the seven gamma ray pulsars indicate a trend
$L_{\gamma}\propto {\dot E}^{1/2}$, and the small additional dependence
on $P^{1/3}$ is still in agreement with this correlation.
If we instead insert the maximum Lorentz factor (\ref{gamma-max}) into
the power formula, the result is
\begin{equation}
L_{\gamma}
=7.9\times 10^{34}\thinspace\frac{\delta^{2}f_{-1}}{\kappa^{6}}
\frac{B_{0,12}^2}{P_{-1}^{4}}
\ {\rm erg\thinspace s^{-1}}\ ,
\end{equation}
and is thus linearly dependent on $\dot E$.

\begin{figure}
\centerline{\psfig{file=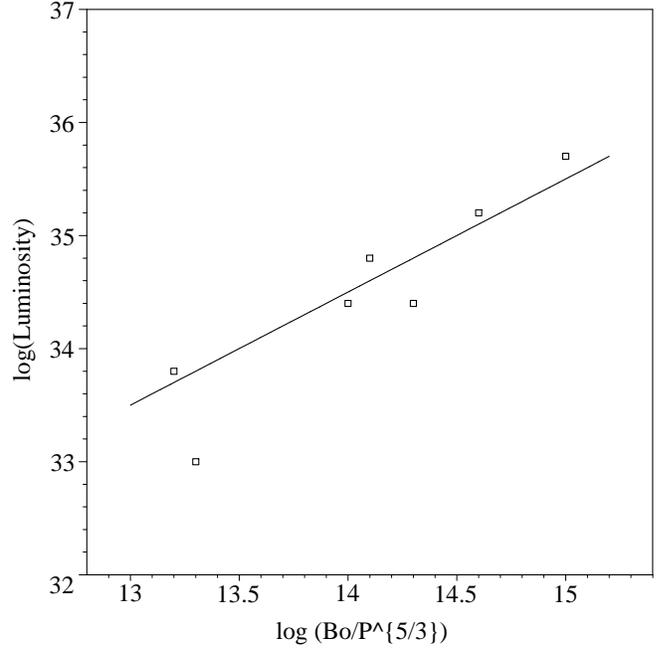,width=8.8cm,clip=} }
\caption{The squares show the observed high
energy luminosities versus the parameter combination
$B_0/P^{5/3}$. The line represents the predicted luminosity at the
break frequency (equation \ref{frequency}), i.e., when
radiative cooling is important.
\label{fig1}}
\end{figure}

\begin{figure}
\centerline{\psfig{file=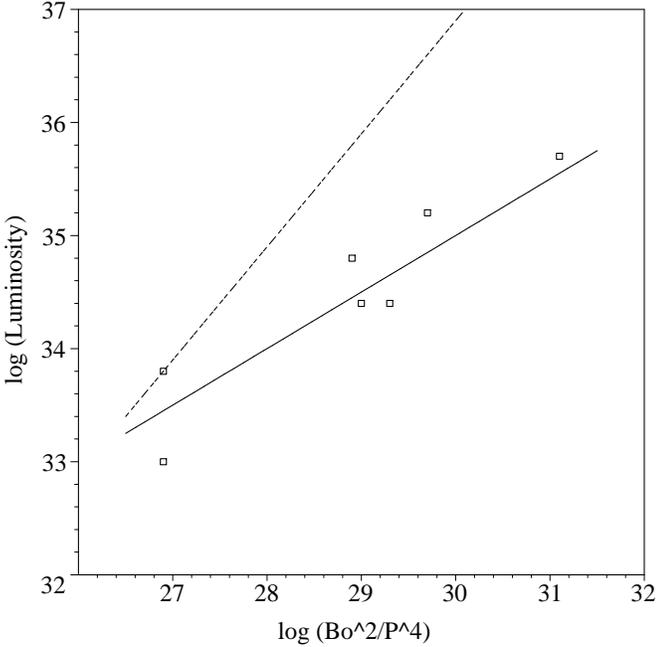,width=8.8cm,clip=} }
\caption{Here the data points are plotted versus
$B_0^2/P^4$. The solid line is given by $L_{\gamma}\propto {\dot
E}^{1/2}\propto B_0/P^2$ and the dotted line represents the
predicted luminosity at the maximum frequency (equation
\ref{numax}), i.e., when radiative cooling is not important.
\label{fig2}}
\end{figure}

In Figures (1) and (2) we compare the model results with
the data points. The latter have been taken from Thompson et al. (1999).
These observed luminosities are integrated fluxes above 1~eV.
It is clearly seen in Figure~(1) that if the maximum of the luminosity
is determined by
the break, a reasonable correlation is found, not only for the slope but also
for the absolute values. However, the observed luminosity of Geminga is
smaller by about a factor of 5, which might be due to the fact that
this pulsar emits 20\% of its spin-down power in gamma rays. This means
that the model curve would give an efficiency of 100\%, which then
causes a significant nonlinear back-reaction on the rotation of the
pulsar and on the
formation of outer gaps, with the effect that the gamma luminosity
of Geminga is reduced below the expectation from the linear model.
Another point worth to note is that the
Crab pulsar emits it maximum spectral power at $\sim 100$ keV.
Therefore the data point of the Crab pulsar that gives the
luminosity in the GeV range is actually somewhat lower than plotted here.
With these remarks in mind our simple model is in good agreement with the data. 
In Figure~(2) the luminosity is plotted versus $B_0^2/P^4$. The solid line
represents the dependence $\log{L_{\gamma}}=20+0.5\thinspace\log(B_0^2/P^4)$,
and the
dotted line is the relation found, if the peak luminosity
would be reached at the cutoff energy. The latter clearly overpredicts
the observed values, especially for the high $B$ -- low $P$ pulsars,
since the slope of the curve is too steep.

\section{Synchrotron Model for the Infrared, Optical and X-Ray Emission
of the Crab pulsar}
 
The Crab pulsar has a continuous spectrum from the optical to X-rays and
$\gamma$-rays with different power-laws (Lyne \& Graham-Smith 1990).
The spectral index $\alpha$, defined as $I_{\nu}\propto
\nu^{-\alpha}$, varies from $\alpha=1.1$ in the $\gamma$-region via
$0.7$ in the hard X-ray region to $0.5$ at soft X-rays (Toor \& Seward 1977) and
zero at optical frequencies (Percival et al. 1993). In the far infrared
region the spectrum is inverted $\alpha=-2$ and cuts off sharply towards
lower frequencies. This behavior is
not accompanied by dramatic pulse profile changes, as one would expect from
saturation or self-absorption effects. Self-absorption should first
influence the peak intensity. The infrared spectrum
measured by Middleditch, Pennypacker, \& Burns (1983) is still not explained.
In the following it is shown that the theory of optically thin
synchrotron radiation at very small pitch angles gives a possible solution
to this problem.

\subsection{Synchrotron Emission at Small Pitch Angles}

The emissivity at very low pitch angles ($\Psi< 1/\gamma$) of a single
particle is given by (Epstein 1973)
\begin{eqnarray}
\epsilon_{\nu}(\theta,\gamma)
={\pi e^2\gamma\Psi^2\nu^3\over{\nu_Bc}}\left[1-{\nu\over{\gamma\nu_B}}+
{\nu^2\over{2\gamma^2\nu_B^2}}\right]\nonumber\\
\times\ \delta\left(\nu-{2\gamma\nu_B\over{1+\theta^2\gamma^2}}\right)
\end{eqnarray}
with $\nu_B=eB/2\pi mc=2.80\times 10^{12}B_6\thinspace {\rm Hz}$.
This formula has to be applied when angles $\theta< 1/\gamma$ are
resolved in the observations, i.e. when the pulse
width $\Delta\phi$ becomes comparable to the emission cone angle of the
particles, $1/\gamma> 2\pi\Delta\phi$.
Since the maximum of the emission is in the forward direction ($\theta=0$)
the emission is dominated by the field lines that point towards the observer
at each phase of the pulse.
When the typical angle of emission $\theta$ is small compared to
the pulse width it is useful to integrate the emissivity
over all angles (averaging):
\begin{eqnarray}
\varepsilon_{\nu}(\gamma)&=&2\pi\int\epsilon_{\nu}\sin\theta\ d\theta\nonumber\\
&=&{2\pi^2e^2\Psi^2\over{c}}\nu\left[1-{\nu\over{\gamma\nu_B}}+
{\nu^2\over{2\gamma^2\nu_B^2}}\right]\ .
\end{eqnarray}

\subsection{Infrared Spectrum}

The emission from a power law distribution
of relativistic particles, $N(\gamma)=N_0\gamma^{-s}$,
in the direction along the magnetic field $\theta=0$ is given by
\begin{eqnarray}
I_{\nu}&\propto&\int\epsilon_{\nu}(0,\gamma) N(\gamma)\thinspace d\gamma\nonumber\\
&\propto&{\pi e^2\Psi^2N_0\nu^3\over{2\nu_B^2c}}\int\delta\left(\gamma-{\nu\over2\nu_B}
\right)\gamma^{1-s}\thinspace d\gamma
\end{eqnarray}
so that
\begin{equation}
I_{\nu}\propto{4N_0\pi e^2\Psi^2\nu_B\over c}\left({\nu\over2\nu_B}\right)^{4-s}\ .
\end{equation}
The spectrum rises very steeply, in fact $I_{\nu}\propto\nu^2$ for $s=2$.
The Lorentz factors in the case of the
Crab then have to be of the order of $\gamma\sim 1/2\pi\Delta\phi\sim 20$,
which is is consistent with the requirement
that $2\gamma\nu_B$ lies in the infrared part of the spectrum.

\subsection{Optical Spectrum}

In case of somewhat higher Lorentz factors the single particle
beam angle becomes smaller than the phase angle of the pulse and the total
(or angle averaged) power spectrum has to be applied.
For simplicity the monochromatic approximation is used:
\begin{equation}
\varepsilon_{\nu}\approx -{dE\over dt}\thinspace\delta(\nu-2\gamma\nu_B)\ .
\end{equation}
This is a good approximation
since the emission is sharply peaked at $2\gamma\nu_B$ and one may just put
all the emission at this frequency.
The result for the optical emission is then
\begin{equation}
I_{\nu}\propto\int\varepsilon_{\nu} N(\gamma)\ d\gamma
={4\pi^2e^2N_0\nu_B^2\Psi^2\over{3c}}\left({\nu\over2\nu_B}\right)^{2-s}\ .
\end{equation}
In the case of $s=2$ this gives the observed flat, $\alpha\approx 0$,
optical/UV spectrum of the Crab pulsar, with $\gamma\sim 10^2$.

\subsection{Soft X-ray Spectrum}

Soft X-ray emission in this model
comes from the particles with larger pitch angles ($\Psi\gg 1/\gamma$),
radiating at a typical frequency $\nu_c={3\over 2}\nu_B\Psi\gamma^2$
The spectrum in this large pitch angle limit is given by
$I_{\nu}\propto \nu^{-(s-1)/2}$. This yields the
observed $I_{\nu}\propto\nu^{-0.5}$ at soft X-rays for $s=2$.

\subsection{Luminosity Estimates for the Optical and X-Ray Emission}

The energy radiated by the pulsar in the optical and in the
X-ray band can be estimated simply as the number of particles times the energy
radiated by one particle $L=Mn_{\rm GJ}VP_{\rm syn}$, where
$M=\gamma_{\rm p}/\gamma$ is the multiplicity of the
relativistic particle density within the volume $V$
as compared to the Goldreich-Julian density $n_{\rm GJ}$.
The volume is estimated as a spherical shell within a fraction $f$ of
the light cylinder radius
$V=4\pi R_{\rm lc}^2\cdot fR_{\rm lc}=1.4\times 10^{30}fP^3\ {\rm cm^{3}}$.
The frequency of the optical emission (at small pitch angles) is given by
\begin{equation}
\nu_{\rm opt}=5.6\times 10^{14}\gamma_{\rm opt,2}B_6\ {\rm Hz}
\end{equation}
and the optical luminosity:
\begin{eqnarray}
L_{\rm opt}=10^{33}\alpha\delta_{-1}f_{-1}\Psi_{\rm p, -3}^{-1}
\Psi_{\rm opt, -2}^2\nu_{\rm opt, 14.7}\nonumber\\
\times\ B_{0, 12.5}^{3/2} P_{-1.5}^{-5/2}\ {\rm erg\thinspace s^{-1}}\ .
\end{eqnarray}
The frequency of the X-ray emission is
\begin{equation}
\nu_{\rm x}=2.8\times 10^{17}\Psi_{\rm x,-1}\gamma_{\rm x,3}^2B_6\ {\rm Hz}
\end{equation}
and the soft X-ray luminosity evaluates to
\begin{eqnarray}
L_{\rm x}=6\times 10^{35}\alpha\delta_{-1}f_{-1}\Psi_{\rm p,-3}^{-1}
\Psi_{\rm x,-1}^{3/2}\nu_{\rm x,17}^{1/2}\nonumber\\
\times\ B_{0,12.5}^{2} P_{-1.5}^{-4}\ {\rm erg\thinspace s^{-1}}\ .
\end{eqnarray}
This can now be compared with the spin-down luminosity $L_{\rm sd}$ of the Crab pulsar
due to magnetic dipole radiation:
\begin{equation}
L_{\rm x}\approx 2\times 10^{-3}\alpha\delta_{-1}f_{-1}
\Psi_{\rm p,-3}^{-1}\Psi_{\rm x,-1}^{3/2}
\nu_{\rm x,17}^{1/2}\thinspace{\dot E}\ .
\end{equation}
Since this relation is now independent of $B_0$ and $P$, it should apply to any pulsar.
Becker \& Tr\"umper (1997) have found such a relation observationally for X-ray
selected pulsars. More details of the model presented here can be found in Crusius-W\"atzel,
Kunzl, \& Lesch (2001).

\section{Discussion and Conclusions}

We have found that curvature radiation cooling results in a universal
spectral break energy in the photon spectrum at $\sim 1$ GeV in gamma
ray pulsars. This break depends
only on geometric factors, but does not depend on other pulsar parameters as
the magnetic field strength or the rotation period. The maximum power is
emitted at this energy. In fact in several of the gamma-ray pulsars
a spectral break or cut-off can be observed at this energy.
A photon spectral index change
of $\Delta\phi=-1$ is expected, when going from lower to higher energies.
This is clearly seen in PSR B1706--44 (Thompson et al., 1996) and
we expect that during the GLAST mission
more pulsars with this spectral feature will be detected.

The high energy luminosity calculated with cooling is proportional
to $B_0/P^{5/3}$, which compares well with the observed values.
Romani (1996) has obtained cutoff energies from a
curvature radiation model without cooling that also
lie in the observed range,
but they depend on the pulsar parameters $B_0$ and $P$,
since they are connected to the
maximum energy of the primaries.
Using the cutoff particle energies for the luminosity estimate we find
that the power output of pulsars in high energy photons
scales as $\propto B_0^2/P^4$, i.e. as a constant fraction of the
spin-down power $\dot E$, which overpredicts the $\gamma$-ray luminosity
for young pulsars. But note that in deriving this result it was
assumed that the density of the emitting particles is of the order
of the Goldreich-Julian density, which might not be true in general.
Harding (1981) estimated the gamma ray luminosity from a polar cap model
by assuming that the power put into particles is completely converted
into radiation and found a scaling $\propto B_0P^{-2}$. In the outer gap
model with a power-law energy distribution of relativistic particles
(due to gap closure), the ``luminosity'' in particles is dominated by
the density at the lowest energy. On the other hand the maximum spectral
power is dominated by the high energy particles with $\gamma=\gamma_{\rm b}$
or $\gamma=\gamma_{\rm max}$. So it is not immediately clear what fraction
of the particle luminosity is converted into radiation.

According to Kanbach (2001) the observed luminosity trend
$L_{\gamma}\propto{\dot E}^{1/2}$ can be caused either
by a proportionality to the number of particles, which
would all be accelerated to the same energy, or by the same particle flow
from different pulsars with correspondingly different acceleration
energies. In our opinion the reality is somewhere in between,
depending on the details of the pair cascade and the gap closure. Only
when the energy (depending on the locally available electric field and
curvature radius) and the density of the relativistic particles,
including its spacial dependence is known,
we can get more insight into the spectrum and the luminosity
of gamma-ray pulsars.

Synchrotron emission of particles with small pitch angles with a single
power-law energy distribution, $N(\gamma)\propto\gamma^{-2}$ reproduces
the key properties of the Crab pulsar spectrum and the
luminosity from the infrared up to soft X-rays.

\begin{acknowledgements}
This work was supported by the Deutsche Forschungsgemeinschaft with the grant
LE 1039/6 (C.-W.).
\end{acknowledgements}


\clearpage

\end{document}